\documentclass[rnote,oldversion,a4paper]{aa}
\usepackage{graphics,txfonts}

\newcommand{\esp}{ESPaDOnS}

\newcommand{\equ}{$\gamma$~Equ}
\newcommand{\hd}{HD\,24712}
\newcommand{\hdalt}{HR\,1217}
\newcommand{\kms}{km\,s$^{-1}$}
\newcommand{\ms}{m\,s$^{-1}$}

\newcommand{\bz}{$\langle B_{\rm z} \rangle$}
\newcommand{\vz}{$\langle v \rangle$}
\newcommand{\dfrac}[2]{\frac{\displaystyle #1}{\displaystyle #2}}

\newcommand{\figps}[1]{\resizebox{\hsize}{!}{\rotatebox{0}{\includegraphics{#1}}}}

\newcommand{\beq}{\begin{equation}}
\newcommand{\eeq}{\end{equation}}

\begin{document}

\title{A high-precision search for magnetic field oscillations \\ 
in the roAp star HD\,24712\thanks{Based on observations obtained at the Canada-France-Hawaii Telescope (CFHT) which is 
operated by the National Research Council of Canada, the Institut National des Sciences de 
l'Univers of the Centre National de la Recherche Scientifique of France, and the University of Hawaii.}}

\author{O. Kochukhov\inst{1} \and G. A. Wade\inst{2}}
 
\offprints{O. Kochukhov, \email{oleg@astro.uu.se}}

\institute{Department of Astronomy and Space Physics, Uppsala University, SE-751 20, Uppsala, Sweden
      \and Department of Physics, Royal Military College of Canada, Kingston, Ontario, K7K 4B4, Canada}

\date{Received 26 December 2006 / Accepted 27 February 2007}

\abstract{
We have obtained a time series of 81 high-cadence circular polarization observations 
of the rapidly oscillating Ap star \hd\ with the new \esp\ spectropolarimeter at CFHT. 
We used the high-S/N, high-resolution Stokes $I$ and $V$ spectra to investigate possible variation 
of the mean longitudinal field over the pulsation cycle in this roAp star. Our multiline magnetic 
field and radial velocity measurements utilized 143 spectral lines of rare-earth elements, 
attaining precision better than 13~G and 19~\ms, respectively. A multiperiodic radial velocity 
variation with an amplitude of 40--136~\ms\ is clearly detected at the known pulsation frequencies
of \hd. At the same time, no evidence for pulsational changes of the magnetic field 
can be found. We derive a $3\sigma$ upper limit of 10~G, or about 1\% of the mean longitudinal
field strength, for magnetic field oscillations in the upper atmosphere of \hd. 
The absence of detectable pulsational variability of the magnetic field provides a valuable constraint
for the interaction between pulsations and magnetic field in roAp stars and is compatible
with the recent predictions of detailed theoretical models of stellar magnetoacoustic oscillations.
}


\maketitle

\section{Introduction}
\label{intro}

Rapidly oscillating magnetic Ap (roAp) stars are cool magnetic chemically peculiar stars
exhibiting a remarkable combination of unusual surface properties. These objects have
strong, organised magnetic fields, most likely of fossil nature, and show photospheric
chemistry deviating far from that of the Sun and other cool, non-magnetic A-type stars.
They show non-uniform  distributions of chemical elements, both laterally across their
surfaces, as well as vertically with height in their atmospheres.  Most remarkably, they
exhibit high-overtone, low-degree, non-radial {\it p-}mode pulsations  with
periods ranging from 6 to 21 min.

Their strong magnetic fields play a central role in the mechanisms responsible for
pulsation mode excitation and selection (Balmforth \cite{BCD01}; Saio \cite{S05}) and in
shaping the frequency spectra of roAp stars (Cunha \cite{C06}). Pulsational perturbations
observed in roAp stars occur in the magnetically-controlled outer stellar layers, which is
why pulsations are aligned with the oblique quasi-dipolar magnetic field  (Kochukhov
\cite{K04}; Saio \cite{S05}) instead of the stellar rotation axis as in all other types of
non-radially pulsating stars.

Recent time-resolved spectroscopic observations of pulsations in roAp stars (Kochukhov \&
Ryabchikova \cite{KR01a,KR01b}; Mkrtichian et al. \cite{MHK03}; Elkin et al. \cite{EKM05};
Kochukhov \cite{K06}) demonstrate that, in addition to the remarkable horizontal geometry,
the magnetoacoustic {\it p-}modes show outstanding vertical structure due to the
combination of the rapid increase of pulsation amplitude with height in the atmosphere and
the short vertical wavelength of the pulsational fluctuations. The general picture of the
atmospheric pulsational behaviour of roAp stars is determined by the propagation of
magnetoacoustic waves through the distinct layers of the chemically stratified atmosphere
(Ryabchikova et al. \cite{RPK02}). Pulsations are weak or non-detectable in the lower
atmosphere probed by lines of light and iron-peak elements. On the other hand, rare-earth
elements (REE) are concentrated in clouds located high above the photosphere, in the layers
where pulsation amplitudes reach up to several \kms.

An important new insight into the complex physics of pulsations in roAp stars can be
achieved by the analysis of interaction between the {\it p-}modes and the magnetic field.
The first search for variations of the mean longitudinal magnetic field, \bz, over the
pulsation cycle was conducted by Hubrig et al. (\cite{HKB04}). These authors used
low-resolution time-resolved circular polarization observations of six roAp stars, and
measured \bz\ using hydrogen lines and unresolved blends of metal lines. Hubrig et al.
(\cite{HKB04}) failed to detect magnetic field variation above the noise level of their
analysis, which was 40--100~G.

Kochukhov et al. (\cite{KRL04}) and Savanov et al. (\cite{SHM06}) obtained high-precision
measurements of the mean magnetic field modulus over the 12-min pulsation period of the
bright roAp star \equ. The resolved, Zeeman-split profile of the \ion{Fe}{ii} 6149.25~\AA\
line was used in both studies. 
Neither of the authors detected any
pulsational fluctuation of the field strength, deriving an
upper limit of 10~G for the possible field variation.

The low-resolution spectropolarimetry and the field modulus measurements using \ion{Fe}{ii}
lines have limited diagnostic capabilities because both types of magnetic field monitoring
techniques probe low atmospheric layers where pulsation amplitudes are small. Taking
guidance from the highly successful radial velocity analyses, an investigation of the
magnetic field oscillations in roAp stars should focus on individual REE lines which show
conspicuous pulsational variations. The first study of this kind has been carried out by
Leone \& Kurtz (\cite{LK03}). They observed several \ion{Nd}{iii} lines in \equ\ using a
circular polarization analyzer and a high-resolution spectrometer. Based on only 18
time-resolved spectra, Leone \& Kurtz (\cite{LK03}) announced the discovery of the
pulsational variation of \bz\ with amplitudes 110--240~G although curiously with discrepant
phases of magnetic maximum for different \ion{Nd}{iii} lines. Subsequently, Kochukhov et
al. (\cite{KRP04}) obtained independent high-resolution polarimetric observations of \equ.
Their analysis combined magnetic signatures of 13  \ion{Nd}{iii} lines and was based on
more than 200 spectra collected over three nights. Kochukhov et al. (\cite{KRP04}) were
unable to detect longitudinal field variability over the pulsation cycle in \equ\ exceeding
40~G. They suggested that the field variation claimed by Leone \& Kurtz (\cite{LK03}) is
spurious and likely results from the neglect of blending of the few REE spectral features
employed in their magnetic measurements.

The contradictory results obtained in the previous searches for rapid magnetic field
oscillations in \equ\ call for an extension of time-resolved spectropolarimetric studies
of REE lines to other roAp stars. We have therefore obtained a new set of time-resolved,
wide wavelength coverage Stokes $I$ and $V$ spectra for the well-known roAp star \hd\
(\hdalt, HIP~18339, DO~Eri). 

\hd\ was identified as a roAp star by Kurtz (\cite{K82}). The recent extensive Whole
Earth Telescope photometric campaign revealed eight pulsation modes with frequencies in the
range of 2.6--2.8 mHz (Kurtz et al. \cite{KCC05}). Ryabchikova et al. (\cite{RLG97})
performed detailed model atmosphere and chemical abundance analysis of \hd. They inferred
$T_{\rm eff}$\,=\,7250~K and $\log g$\,=\,4.3, and found that the atmosphere of this roAp star
is rich in REEs but relatively poor in the iron-peak elements. 
In their study of the evolutionary state of magnetic chemically peculiar stars
Kochukhov \& Bagnulo (\cite{KB06}) derived the following fundamental characteristics
of \hd: mass $M$\,=\,1.55$~M_\odot$, luminosity $L$\,=\,7.4$~L_\odot$ and an age $\approx$\,$10^9$~yr, 
corresponding to about 50\% of the main sequence lifetime.
Bagnulo et al. (\cite{BLL95})
demonstrated that a dipolar magnetic field topology with the polar strength $B_{\rm
p}$\,=\,3.9~kG and the angle $\beta$\,=\,150\degr\ between the rotation and magnetic axes
provides a good fit to both the longitudinal field curve of \hd\ and to the broad-band
linear polarization measurements. Combining magnetic field and line strength observations of
\hd\ obtained over more than three decades, Ryabchikova et al. (\cite{RWA05}) determined a
rotation period of 12.45877~d.

\section{Observations and data reduction}
\label{obs}

Eighty-one new Stokes $V$ spectra of \hd\ were obtained on 9 January 2006 UT using the
\esp\ spectropolarimeter at the Canada-France-Hawaii Telescope. 
Observations covered roughly 3 hours, starting at HJD\,=\,2453744.70540. According to
the ephemeris of Ryabchikova et al. (\cite{RWA05}), our time-resolved monitoring of \hd\
corresponds to the rotation phase 0.9, which is close to the positive extremum of the 
longitudinal magnetic field variation and to the phase of maximum non-radial oscillation
amplitude. Observations were obtained
in polarimetric mode, with a nominal resolving power $\lambda/\Delta\lambda=65000$,
and cover the spectral region 3700--10500~\AA. The details of the \esp\ are
reported by Donati et al. (in preparation)\footnote{For up-to-date information, visit
http://www.ast.obs-mip.fr/projets/espadons/espadons.html.}.
In particular, both results of the commissioning tests and our own velocity measurements
using telluric absorption lines in the spectra of \hd\ indicate that the typical relative
spectral stability of the instrument is better than 10--15~\ms\ for the frequency range 
corresponding to roAp pulsations.

The 81 Stokes $V$ spectra of \hd\ were computed from 162 individual circular polarization
exposures. Between subsequent exposures, the Fresnel rhomb quarter-wave retarder was
rotated by $90\degr$ (i.e. alternately to position angles of $+45\degr$ and $-45\degr$) in
order to interchange the positions of the ordinary and extraordinary beams on the CCD. The
data were reduced using the Libre-ESpRIT package; each subsequent pair of CCD frames was
combined into a single Stokes $V$ spectrum with typical peak S/N of 115. Examination of
the diagnostic null spectrum (Donati et al. \cite{DSC97}) indicates that the beam-switching
technique has removed all spurious contributions to the Stokes $V$ spectrum to below the
noise level.
We emphasize that this is the first time that this high-precision spectropolarimetric 
observing method has been applied to monitor rapid magnetic field variations of a roAp star.

Individual circularly polarized exposures were of 20 seconds duration, and the readout using
the CCD's fast readout mode required 25 seconds. Taking into account time required for
repositioning of the waveplate between exposures, the average time required for acquisition
of a single Stokes $V$ spectrum was typically 1.25 minutes. This effective time
resolution of the time series is significant in comparison with the $P$\,=\,6.0--6.4~min
pulsational variation of \hd. 
Our numerical modelling shows that the pulsational amplitude inferred from such
a time series is underestimated by 12\% relative to observations with a much higher time
resolution. The treatment of this effect must be included in the time-series analysis.

\section{Time-series analysis of radial velocity and magnetic field}
\label{rvbz}

\begin{figure*}[!th]
\figps{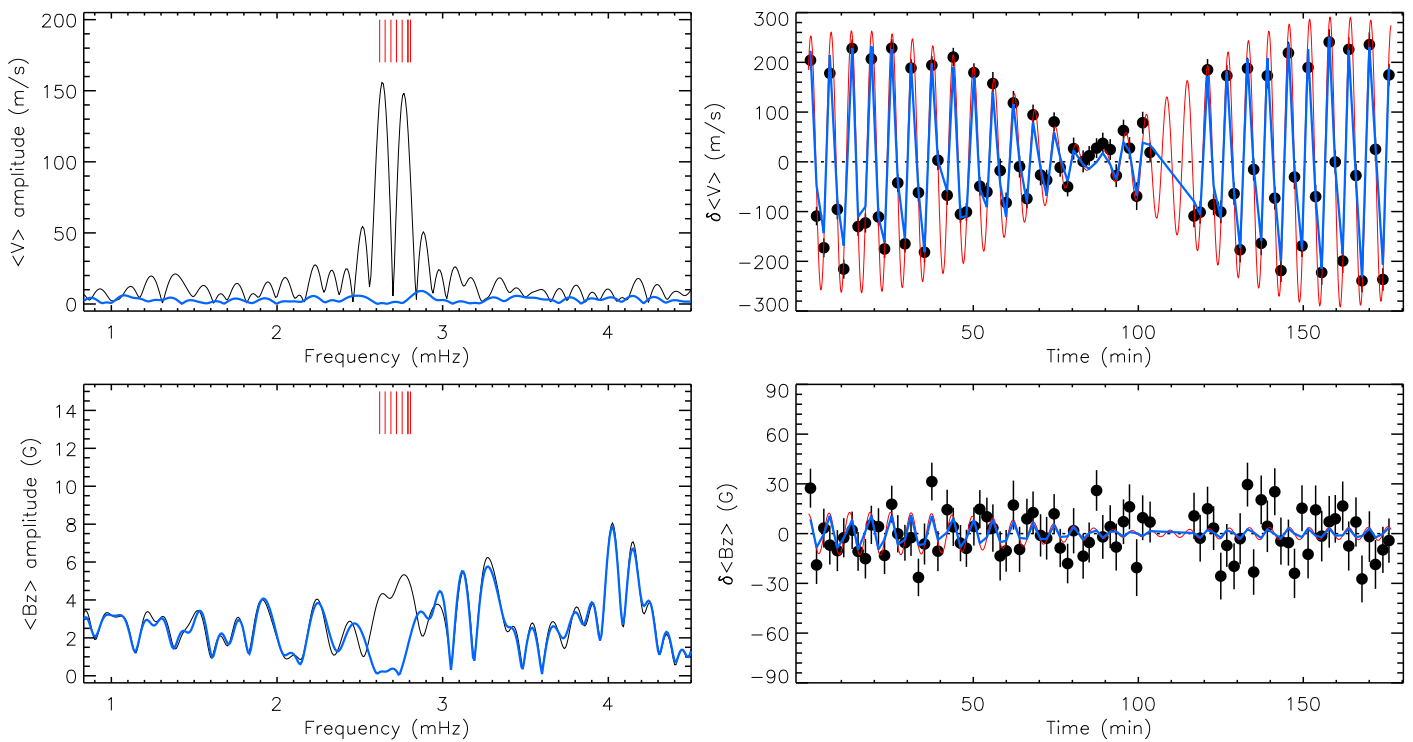}
\caption{Pulsational variation of radial velocity ({\it upper panels}) and longitudinal field 
({\it lower panels}) for the REE lines in \hd. The left panels show the amplitude spectra of RV 
and \bz\ ({\it thin line}) as well as Fourier transforms of the residual variation 
({\it thick line}) obtained by subtracting least-squares sinusoidal fit from the original time
series. The vertical bars 
show pulsation frequencies of \hd\ determined in the photometric study by Kurtz et al. (\cite{KCC05}). 
The right panels illustrate velocity and magnetic field variation in the time domain. 
Individual REE and $\delta$\bz\ measurements ({\it symbols}) are compared with the best fit 
three-frequency model ({\it thick line}), which takes into account data sampling and 
phase smearing (see text). The thin line shows the same oscillation model neglecting these 
effects.}
\label{fig2}
\end{figure*}

Ryabchikova et al. (\cite{RSW06}) have recently presented a detailed study of the spectroscopic
pulsational behaviour of \hd\ using high-resolution \'echelle spectra acquired at several large telescopes.
This analysis demonstrated that, similar to other roAp stars, the pulsational radial velocity (RV)
variation can be clearly detected only for REE lines and, marginally, for the cores of a few very strong
lines of light elements (hydrogen Balmer lines, \ion{Ca}{ii} 3933~\AA). Taking these results into
account, we have investigated the pulsational variation of the Stokes $I$ and $V$ profiles for REE
lines. An initial list of the central wavelengths and Zeeman splitting parameters was compiled for
more than 200 lines based on the information from Ryabchikova et al. (\cite{RRK06,RSW06}), and the VALD
(Kupka et al. \cite{KPR99}) and DREAM (Bi\'emont et al. \cite{BPQ99}) databases. This list is
dominated by \ion{Pr}{iii}, \ion{Nd}{ii} and \ion{Nd}{iii} lines, but also includes 
\ion{Y}{ii}, \ion{La}{ii}, \ion{Ce}{ii}, \ion{Pr}{ii}, \ion{Sm}{ii}, \ion{Eu}{ii}, \ion{Gd}{ii}, 
\ion{Tb}{iii}, \ion{Dy}{ii}, \ion{Dy}{iii}, and \ion{Er}{iii} spectral features. The Stokes $I$ and
$V$ profile of each line was examined to remove lines affected by blending. Radial velocities were
determined for the remaining lines with the center-of-gravity method:
\beq
\langle v \rangle = \frac{c}{\lambda_0} \left( 
\frac{\int \lambda (I_{\rm c}-I) {\rm d}\lambda}{\int (I_{\rm c}-I) {\rm d}\lambda} - 
\lambda_0 \right),
\label{eq0}
\eeq
where $\lambda_0$ is the laboratory wavelength, $I$ and $I_{\rm c}$ correspond to the line and
continuum intensity, respectively, and $c$ is the speed of light.

The mean longitudinal magnetic field was estimated by integrating the Stokes $V$ profiles of REE
lines:
\beq
\langle B_{\rm z} \rangle = 
-\dfrac{4\pi m_{\rm e}c}{e}
\frac{\int (v-\langle v \rangle) V {\rm d}v}{\lambda_0 z \int (I_{\rm c}-I) {\rm d}v}, 
\label{eq1}
\eeq
where $z$ is the effective Land\'e factor, $m_{\rm e}$ is the 
electron mass, $e$ is the electron charge, and $v$ is the velocity coordinate.
The errors of \vz\ and \bz\ were derived 
applying the standard error propagation rules to Eqs.~(\ref{eq0}) and (\ref{eq1}), and using
the formal uncertainties of the Stokes $I$ and $V$ spectra provided by the Libre-ESpRIT spectral
extraction package.

Preliminary time-series analysis was carried out for each measured transition to exclude lines
showing no RV oscillation signal above $2\sigma$ in the 2.4--3.0~mHz frequency interval to which all
known pulsation frequencies of \hd\ are confined (Kurtz et al. \cite{KCC05}). 
Four \ion{Tb}{iii} were also excluded from consideration because their pulsational radial velocity
curves show a large phase shift with respect to the variation of other REE ions (Ryabchikova et al. \cite{RSW06}).
This left us with 139 REE lines. Due to overlap of the blue \'echelle orders in the \esp\ spectra, 
many lines could be measured twice, increasing the total number of spectral features included in the
final sample to 168.

\begin{table*}[!th]
\caption{Results of the multifrequency least-squares analysis of the variation of radial velocity and longitudinal magnetic field 
for REE lines in \hd. For each group of lines we list the number of measured spectral features,
followed by the number of unique lines. Amplitudes are given in \ms\ and G for \vz\ and 
\bz, respectively, while phases are measured in fractions of pulsation period.}
\label{tbl1}
\begin{center}
\begin{tabular}{crcrc}
\hline
\hline
$P$    & \multicolumn{2}{c}{Radial velocity} & \multicolumn{2}{c}{Magnetic field} \\
 (min) & $A$ (\ms) & $\varphi$ & $A$ (G)~~ & $\varphi$ \\
\hline
\multicolumn{5}{c}{168(139) REE lines} \\
 6.049 & $138.7\pm3.9$ & $0.850\pm0.005$ &  $4.9\pm2.4$ & $0.857\pm0.076$ \\
 6.282 & $113.9\pm6.2$ & $0.861\pm0.009$ &  $6.6\pm3.8$ & $0.143\pm0.098$ \\
 6.362 &  $40.9\pm6.3$ & $0.120\pm0.024$ &  $5.1\pm3.9$ & $0.944\pm0.118$ \\
\multicolumn{5}{c}{56(47) \ion{Nd}{ii} lines} \\
 6.049 & $189.6\pm6.0$ & $0.936\pm0.005$ &  $5.2\pm5.6$ & $0.399\pm0.174$ \\
 6.282 & $132.0\pm9.5$ & $0.971\pm0.012$ & $19.9\pm9.0$ & $0.317\pm0.072$ \\
 6.362 & $ 41.3\pm9.7$ & $0.139\pm0.037$ &  $8.0\pm9.1$ & $0.808\pm0.184$ \\
\multicolumn{5}{c}{44(34) \ion{Nd}{iii} lines} \\
 6.049 & $152.0\pm4.3$ & $0.842\pm0.005$ &  $4.8\pm3.2$ & $0.944\pm0.106$ \\ 
 6.282 & $138.6\pm6.9$ & $0.851\pm0.008$ &  $7.4\pm5.1$ & $0.196\pm0.108$ \\ 
 6.362 &  $43.7\pm7.1$ & $0.148\pm0.026$ & $10.3\pm5.2$ & $0.037\pm0.079$ \\ 
\multicolumn{5}{c}{17(16) \ion{Pr}{iii} lines} \\
 6.049 & $179.5\pm~~8.6$& $0.662\pm0.007$ & $16.5\pm~~7.6$& $0.859\pm0.075$ \\
 6.282 & $173.7\pm13.5$ & $0.663\pm0.012$ & $18.4\pm12.1$ & $0.902\pm0.105$ \\ 
 6.362 &  $73.8\pm13.8$ & $0.975\pm0.030$ &  $8.7\pm12.2$ & $0.754\pm0.227$ \\ 
\multicolumn{5}{c}{15(12) \ion{Sm}{ii} lines} \\
 6.049 & $213.5\pm10.5$ & $0.923\pm0.008$ & $10.5\pm~~9.2$& $0.829\pm0.143$ \\
 6.282 & $191.4\pm16.9$ & $0.966\pm0.014$ & $30.0\pm14.9$ & $0.678\pm0.078$ \\
 6.362 &  $26.3\pm17.2$ & $0.419\pm0.104$ & $18.4\pm15.1$ & $0.474\pm0.130$ \\
\hline
\end{tabular}
\end{center}
\end{table*}

We have constructed a high-precision measure of the pulsational changes of RV and magnetic field by
calculating the weighted average of the deviation of \vz\ and \bz\ from their mean values:
\beq
\overline{\delta S} = \frac{\sum_{i=1}^{N} \left[S_i - (a_i + b_i t)\right]/\sigma_i^2(S)}
{\sum_{i=1}^{N} 1/\sigma_i^2(S)},
\eeq
where $N=168$, $S$ corresponds to $\langle v \rangle$ or $\langle B_{\rm z} \rangle$, $\sigma_i(S)$ are the 
respective errors and $a_i$, $b_i$ represent coefficients of a linear least-squares fit to the full 
timeseries describing the mean and linear drift of each quantity. 

The amplitude spectrum and time dependence of the resulting average $\delta$\vz\ and $\delta$\bz\ are
illustrated in Fig.~\ref{fig2}. We have unambiguously detected pulsational variation of RV at the
level of $\approx$\,150~\ms. Pronounced modulation of the RV amplitude due to beating of the
oscillations with $P$\,$\approx$\,6.05 and 6.30~min is evident in our data. The first period coincides with the
frequency $\nu_5$ (2755~$\mu$Hz, 6.049~min) of Kurtz et al. (\cite{KCC05}), whereas the second one
lies between $\nu_1$ (2620~$\mu$Hz, 6.362~min) and $\nu_2$ (2653~$\mu$Hz, 6.282~min). Given that the
frequency spectrum of \hd\ is well known and remains constant over many years, we assume that the
{\it p-}mode oscillations detected in our CFHT spectra are caused by the superposition of the three 
pulsation periods corresponding to $\nu_1$, $\nu_2$ and $\nu_5$ of Kurtz et al. (\cite{KCC05}). 
We keep these periods fixed in the following time-series analysis.

Figure~\ref{fig2} shows that there is no evidence of the \bz\ variation with any of the known
pulsation frequencies of \hd, nor any other frequency up to the Nyquist limit of our data. The
highest $\delta$\bz\ peaks reach 6--8~G, but none is statistically significant.

We have used a least-squares analysis in the time domain to determine RV amplitudes and to
infer accurate upper limits for the magnetic field variation. 
For high-cadence data in which the time sampling is much shorter than the pulsation period, the expression
\beq
\overline{\delta S} = \sum_{i=1}^3 A_i \cos{\left(\frac{2\pi t}{P_i}+2\pi\varphi_i\right)}
\label{eq3}
\eeq
(where $A_i$ and $\varphi_i$ are the amplitude and phase of oscillation with period $P_i$) is 
suitable to approximate the variability of pulsating stars. However, the phase smearing caused by 
the longer effective time sampling used in this study is non-negligible. We have explicitly taken the data
sampling effects into account by integrating and averaging Eq.~(\ref{eq3}) for the time intervals
corresponding to the individual polarimetric subexposures. 
Applying trivial trigonometric relations, one can obtain the following expression:
\beq
\begin{array}{rl}
\overline{\delta S} = & \displaystyle\sum_{i=1}^3 \dfrac{A_i P_i}{\pi\Delta t} 
\sin{\left(\dfrac{\pi \Delta t}{P_i}\right)}
\cos{\left(\dfrac{\pi t^{+45}+\pi t^{-45}}{P_i}+2\pi\varphi_i\right)} \\
 & \times\cos{\left(\dfrac{\pi t^{+45}-\pi t^{-45}}{P_i}\right)},
\end{array}
\label{eq4}
\eeq
where $t^{+45}$ and $t^{-45}$ are mid times of the subexposures with different retarder orientation
and $\Delta t=20$~s is the exposure time. 

Results obtained by fitting Eq.~(\ref{eq4}) to the \vz\ and \bz\ variations are summarized in
Table~\ref{tbl1} and illustrated in Fig.~\ref{fig2}. The RV pulsations have amplitudes 40--136~\ms\
and are detected at the 6--36$\sigma$ level. The rms deviation between observations and the
least-squares fit indicates an uncertainty of 19~\ms\ for a single multiline RV measurement.
The corresponding precision of the multiline magnetic measurements is 13~G. Fig.~\ref{fig2} shows that the amplitude 
of the residual oscillations does not exceed 5--10~\ms\ after prewhitening observations with the 
three-frequency model. At the same time, the formal \bz\ amplitudes are 5--6~G, detected at the 
2$\sigma$ confidence level or less. Therefore, we conclude that \hd\ exhibits no pulsational
variation of the disk-averaged line of sight magnetic field component. The 3$\sigma$ upper limit for
pulsational field changes is 7--12~G, depending on the pulsation mode considered.

We have examined separately the average RV and \bz\ for several groups of REE ions. Table~\ref{tbl1}
presents results of this analysis for the four groups of REE species with $>$\,10 unique lines each
(\ion{Pr}{iii}, \ion{Nd}{ii}, \ion{Nd}{iii}, \ion{Sm}{ii}). The time-series analysis is less precise
for individual ions, but it generally confirms results obtained with all REE lines. In particular,
none of the investigated REE species shows statistically significant pulsational changes of \bz,
although each of them shows RV pulsations at $>$20$\sigma$ level.

\begin{figure}[!th]
\figps{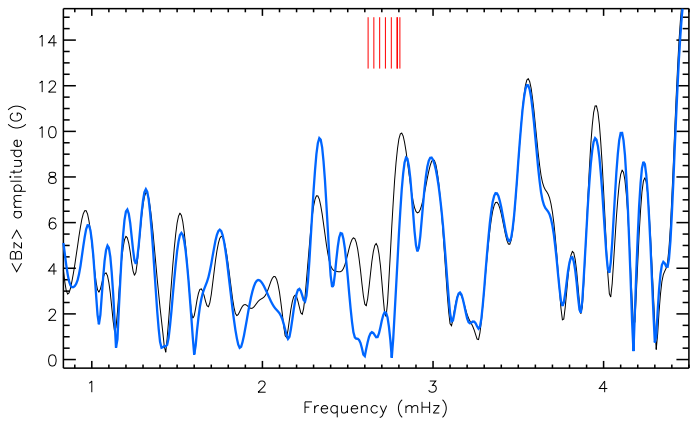}
\caption{Amplitude spectra of the LSD \bz\ measurements for \hd. Fourier transform
of the original \bz\ time series ({\it thin line}) is compared with the residual amplitude
spectrum ({\it thick line}). The vertical bars show photometric pulsation frequencies of \hd.}
\label{fig3}
\end{figure}

The REE line analysis was complemented by the study of \bz\ measured for the 
Least Squares Deconvolved Stokes $I$ and $V$ spectra (LSD profiles, see Donati et al.
\cite{DSC97}). This diagnostic is based upon strong lines of light and iron-peak elements and,
therefore, probes conditions in the lower part of \hd\ atmosphere. The amplitude spectrum of the
LSD \bz\ measurements is presented in Fig.~\ref{fig3}. No variation above 13--21~G (3$\sigma$
limits) is present at the frequencies detected in the REE line radial velocities. The average
longitudinal field obtained with the LSD technique is $1022\pm30$~G, which is in good agreement
with the known rotational modulation of \bz\ in \hd\ (Ryabchikova et al. \cite{RWA05,RSW06}).

\section{Discussion}
\label{disc}

We have obtained the first time-resolved high-resolution circular polarization observations of the
roAp star \hd. Our aim was to search for the variation of the stellar magnetic field with the pulsation
cycle. Unlike all previous time-resolved spectropolarimetric studies of roAp stars (Hubrig et al.
\cite{HKB04}; Leone \& Kurtz \cite{LK03}; Kochukhov et al. \cite{KRP04}), we have employed the
beam-switching technique to minimize spurious polarization. The resulting phase smearing was
accounted for in the time-series analysis. To boost precision further, we have combined oscillation
signal in 139 REE spectral lines, thus significantly improving the sensitivity to the magnetic field
variation in comparison with previous studies. Variability with the known pulsation frequencies was
found for the RV obtained from averaged REE lines as well as for individual ions. 
However, no magnetic field variability above 7--12~G could be detected.
Since our observations were obtained at the rotation phase of magnetic maximum of
\hd, when \bz\,\,$\approx$\,1~kG, we infer the upper limit of the relative pulsational changes of the 
field strength: $\delta B/B\la10^{-2}$.

The null result of our search for the pulsational magnetic field variation in \hd\ agrees with the
2--4\% $\delta B/B$ upper limit obtained by Kochukhov et al. (\cite{KRP04}) for \equ. Our study thus
strengthens the view that no significant field variability in roAp stars should be expected on
pulsational timescales. 

Hubrig et al. (\cite{HKB04}) published a preliminary theoretical estimate of the expected pulsational
modulation of \bz\ in roAp stars. These authors claimed that field variations at the level of
1\% to 14\% for different roAp stars may be expected. For the pulsation period and the radial
velocity amplitude observed in the upper atmosphere of \hd, magnetic changes should be about 1\% of \bz\ or 10~G. 
This level of the magnetic field variations is marginally inconsistent with our conservative
upper limit of 7~G derived for the highest-amplitude pulsation mode. 
At the same time, Hubrig et al. (\cite{HKB04})
predict that the rapid magnetic variations of \equ\ should reach 10\% of the field strength, which is 
clearly ruled out by the observational results of Kochukhov et al. (\cite{KRP04}).

Saio (\cite{S05}) has pointed out that in their estimate of the magnetic field variation Hubrig et
al. (\cite{HKB04}) have erroneously disregarded the effect of the horizontal component of the
pulsational displacement. Based on the detailed non-adiabatic theoretical model of the
magnetoacoustic oscillations in roAp stars, Saio (\cite{S05}) suggested that the amplitude of the
magnetic field variation should not exceed $\delta B/B$\,$\sim$\,$10^{-5}$ or $\sim$\,0.01~G for a star
with a kG-strength magnetic field. This theoretical prediction is confirmed by our spectropolarimetric
monitoring of \hd\ and by the time-resolved magnetic observations of \equ\ presented
by Kochukhov et al. (\cite{KRP04}).

\end{document}